\documentclass[twocolumn,showpacs,preprintnumbers,amsmath,amssymb]{revtex4}

\usepackage{graphicx}
\usepackage{dcolumn}
\usepackage{bm}

\begin{document}

\preprint{APS/123-QED}

\title{Parameter-free extraction of Thin-Film Dielectric Constants from Scanning Near Field Microwave Microscope Measurements
}
\author{Shuogang Huang$^{1}$, H. M. Christen$^{2}$, and M. E. Reeves$^{1,3}$}
\address{$^{1}$Physics Department, The George Washington University, Washington, DC 20052}
\address{$^{2}$Solid State Division, Oak Ridge National Laboratory}
\address{$^{3}$Naval Research Laboratory, Washington, DC 20375}


\date{\today}

\begin{abstract}

We present a method for extracting high-spatial resolution dielectric constant data at microwave frequencies. A
scanning near field microwave microscope probes a sample and acquires data in the form of the frequency
and quality factor shifts of a resonant cavity coupled to the sample. The approach reported here is to calculate the electromagnetic fields by the finite element method in both static and time-dependent modes.  Cavity perturbation theory connects the measured frequency shifts to changes in the computed energy stored in the
electromagnetic field. In this way, the complex permittivity
of the sample is found. Of particular interest are thin-film materials, for which a method is reported here to determine the dielectric constant without the need to use any fitting parameters.

\end{abstract}

\maketitle

\textbf{I. INTRODUCTION}

The scanning near-field microwave microscope is a novel instrument for measuring thin film dielectric properties with both high precision and high spatial resolution.  In this technique, a sharpened probe extending from a resonant cavity is coupled electromagnetically to a sample brought into close proximity.
In the geometry used by our experiment, the fields of interest are well described by the near field approximation, where the spatial resolution is mainly determined by
the tip-sample spacing and the shape of the probe tip, rather than by the wavelength $\lambda$. The challenge of this method is to convert the frequency shift to the sample's permittivity. Several groups have followed an approach of modeling the cavity-sample interaction by an image charge,
located in the sample, which interacts with another image charges located inside the spherical tip.\cite{Gao98}  A further refinement of this approach is to model the tip as a perfectly conducting sphere, and then rely on numerical calculations to compute the frequency shift of bulk or thin film dielectric samples.\cite{Lee01} An alternative approach is to model the tip-sample interaction as an extension of a transmission line.\cite{Steinhauer99, Tselev03} Often these approaches fall short of an exact description of the actual experimental configuration and in these cases, the model must be parameterized by measurements against standard samples in order to compensate for these discrepancies.  In contrast, this paper reports the development of an approach to more accurately model the tip-sample interaction and by so doing, to determine the complex permittivity of dielectric materials at microwave frequencies.  Perturbation theory provides a framework, by which to convert the calculated electric field distribution in the
near-field zone of the microwave probe into the frequency shift of the resonant structure from which the field emanates.  By comparing the calculated to the measured shifts, the dielectric constant is determined with good precision and with high spatial resolution for bulk and thin-film samples.  For bulk samples, the method described does require calibration to account for the unknown energy stored in the cavity, but for this films, no calibration is needed.

In specific, this paper reports a method to extract the dielectric properties of bulk
and thin-film samples from SNMM data that are measured frequency shifts of a $\lambda$/4 cavity with a closed end.
A finely sharpened tungsten wire protrudes from the bottom of the cavity and the sample is raised up into contact
with the tip.  The cavity's resonant frequency and quality factor changes as a function of tip-sample
separation. The field in the vicinity of the tip is computed and the modeled response of he cavity to the introduction of the smple
is compared to the measured quantity.  There are a number of publications on the
experimental method used here \cite{Gao98, Our00} and so the apparatus will not be described in any more detail.
This paper will instead focus instead on extracting the dielectric properties from the cavity's response to the
perturbation of its electromagnetic fields by a sample.  Several aspects will be described: the accurate
measurement of bulk dielectric constants over a wide range of dielectric constants (after a single calibration),
the simulation of the total loss to compute the imaginary component of the bulk dielectric constant, and the
measurement of the dielectric constant of thin-films without the need for calibrations (which are difficult due
to a lack of thin-film standards).  To this end, the paper will be divided into seven parts:
\begin{itemize}
\item Theory and simulation: assumptions and approximations.
\item Geometry and calculation of the static field.
\item Extraction of the calibration constant for bulk measurements.
\item Parameter-free extraction of thin-film dielectric constants.
\item Determination of the imaginary part of the dielectric constant.
\item Experimental measurements of bulk and thin-film samples.
\item Discussion of polarization, sensitivity, and spatial resolution.
\end{itemize}

\textbf{II. Theory and simulation:assumptions and approximations}

In rare cases, the cavity's response to a dielectric perturbation can be calculated in closed form, but more often, simulations and approximations must be used.  To this end, the cavity perturbation method provides a useful framework for the conversion of frequency shifts to a sample's dielectric constant.  The major assumption is that the eigenmodes of the
microwave field in the resonant structure (cavity + tip + surroundings) are unchanged from those of the unperturbed cavity.  Then, the resonant
frequency shift of the cavity can be related to the change in the permittivity ($\Delta\epsilon$), or permeability
($\Delta\mu$).

\begin{equation}\label{equ1:1}
\frac{f-f_{0}}{f}=\frac{-\int\int\int_{V_{0}}(\Delta\epsilon\vec{E}\cdot\vec{E}^{\ast}_{0}+\Delta\mu\vec{H}\cdot\vec{H}^{\ast}_{0})dv}{\int\int\int_{V_{0}}(\epsilon\vec{E}\cdot\vec{E}^{\ast}_{0}+\mu\vec{H}\cdot\vec{H}^{\ast}_{0})dv}
\end{equation}

Here, the subscript 0 represents the
unperturbed case \cite{Pozar}.  To make use of Eq. 1, a further assumption is made that the fields on the inside of the resonant cavity are not changed by the sample, but only the fields in the region of the tip are significantly altered.  This is exemplified by the 1/1000 shift in the resonant frequency of the cavity upon sample introduction (See Fig. 7).  In total, E$_{0}$ = E$_{cavity}$ + E$_{tip}$ + E$_{surrounding}$.  All three fields contribute to the stored energy of the structure.  However, when the sample is introduced, only the changes in E$_{tip}$ + E$_{surrounding}$ need to be considered.

From a practical viewpoint, the calculation proceeds by first computing the electromagnetic fields in the vicinity of a bare tip by numerically solving Poisson's equation using a commercial PDE package, FEMLAB.  The computation is repeated with a geometry, which includes the sample, and because the change in permittivity only occurs within the small volume occupied by the sample, this region is calculated with a dense mesh. The calculated fields from both simulations are exported, and post-processed (using Matlab) to numerically integrate the scalar product of the fields over the volume of the sample.  In this way, the  numerator on the right side of Eq. \ref{equ1:1} is computed.  For nonmagnetic samples, $\Delta\mu=0$, and the perturbed energy can be simplified to be:

\begin{equation}\label{equ212}
    E_{pert}=-\int\int\int_{V_{0}}\Delta\epsilon\vec{E}\cdot\vec{E}^{\ast}_{0}dv.
\end{equation}

\textbf{III. Geometry and calculation of the static field}

\begin{figure}
  \centering
  \includegraphics[width=8cm]{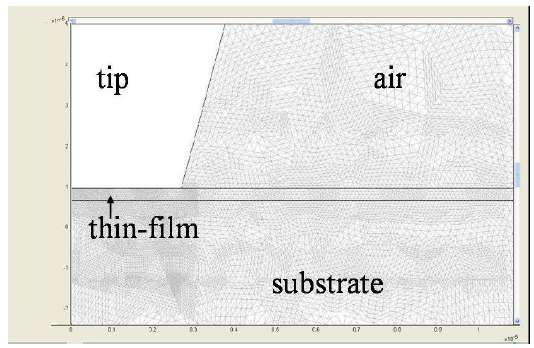}\\
  \caption{Schematic mesh of a dielectric sample in contact with a tip. The tip shown here has a 5$\mu$m diameter flat spot on its end. The thin film is about 400nm thick. The mesh is refined twice in the area where the tip and sample come together.}\label{fig1:1}
\end{figure}

In more detail, the fields are calculated from a FEMLAB geometry, like the one in Fig \ref{fig1:1}.  The high-conductivity, tungsten tip is represented by an equal-potential line, $\phi=1V$. After several measurements, the tip is found to be slightly flattened, as shown in Fig.  \ref{fig1:1}, which qualitatively changes the electric field.  For this reason, the exact tip shape is determined by scanning electron microscopy (SEM) after the experiment.  As can be seen from the Figure, the geometry is axially symmetric, which allows a simplified and more efficient calculation in cylindrical coordinates.

\begin{figure}
  \centering
  \includegraphics[width=8cm]{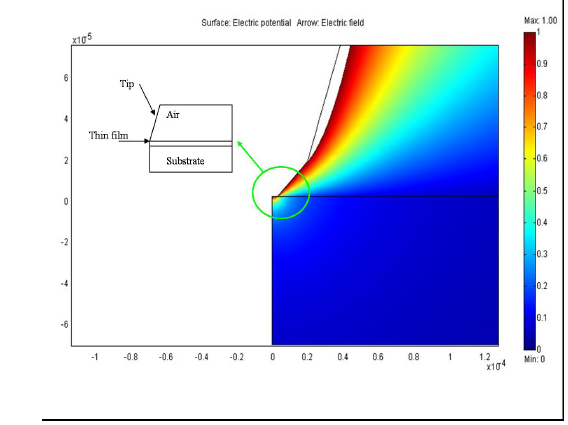}\\
  \caption{The computed, near-field electric potential, static case. The inset is a magnified view of contact area. The electric potential is strongest at the very end of the tip, which results in enhanced sensitivity to the sample below it.}\label{fig1:3}
\end{figure}

The dielectric material under study is placed beneath the tip. Fig. \ref{fig1:1} depicts the `soft contact', the
point at which the tip barely touches the sample. To be consistent with the experiment, a ground plane is added
below the sample, the dielectric constants in all domains are initially set to one (to simulate the case without
a sample), and a mesh is generated. In order to obtain accurate results, the mesh near the tip end is refined
several times, and the near-field electric potential is found by solving the Poisson's Equation.  See Fig.
\ref{fig1:3} and its inset for a magnified view of the tip-film-substrate geometry. The Figure shows that a strong
electric field is confined to a small region (approximately, 1$\mu$m$\times$1$\mu$m) close to the tip end. The
gradient of the potential gives the solution for the the unperturbed field E$_{0}$, the values of which are
exported to Matlab in the form of a FEM structure; `fem1'.

By changing both thin film and substrate domains to the sample's dielectric constant and redoing the simulation
without changing the mesh, the perturbed field for a bulk sample or substrate is calculated. The values of this
solution over the entire geometry are exported as another FEM structure; `fem2'. Proceeding further, the
integral of the inner product of these two vectors over the entire sample volume is calculated to compute
E$_{pert}$.
\\
\textbf{IV. Extraction of the calibration constant for bulk measurements}
\\
The denominator in Eq. \ref{equ1:1} is the total energy of the cavity and its surrounding space, which can be set
equal to a constant $\alpha$, by the approximations already discussed, and Eq. \ref{equ1:1} is rewritten,
\begin{equation}\label{equ213}
    \frac{\Delta{f}}{f}=\frac{E_{pert}}{\alpha}.
\end{equation}

\begin{figure}
  \centering
  \includegraphics[width=8cm]{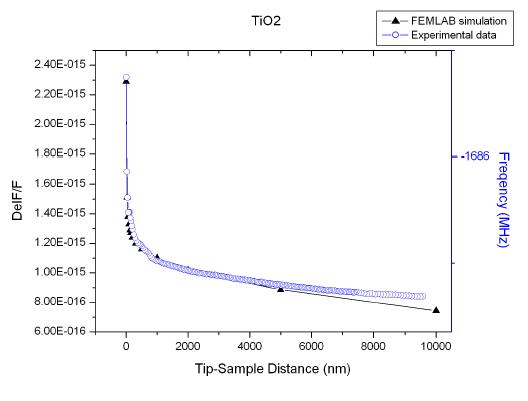}\\
  \caption{Calibration used to find the scaling factor $\alpha$. The experimental data are scaled to lie on top of the with the simulation data, which end up overlapping well when the tip-sample distance is less than 5$\mu$m.  The scale factor, found by calibating against a medium-high permittivity sample like TiO$_2$, is used to successfully measure samples whose permittivities range from 3 to 300.}\label{fig1:4}
\end{figure}

It is impractical to calculate the $\alpha$ numerically, as this would require an accuracy greater than 1ppm for the
integration over the entire space (sample + cavity).  On the other hand, this is hardly necessary, since $\alpha$ may be treated as a parameter found by scaling the calculated energy shift \textit{vs.} tip sample distance curve so that it lies on top of the measured data, as shown in Fig.  \ref{fig1:4}.

This strategy, to compute the calibration constant, $\alpha$, from the dependence of $\Delta$f/f on the distance
between the tip and a calibration standard, is used only for bulk samples.  An example is shown in Fig.
\ref{fig1:4} where E$_{pert}$ and the resonant frequency are plotted as a function of distance between the tip
and a sample of bulk TiO$_2$.  The open circles are the measured data and the triangles are the calculated energy
shifts. The data are obtained at 10 $\mu$m, 5 $\mu$m, 2 $\mu$m, 1 $\mu$m, 500 nm, 300 nm, 200 nm, 180 nm, 150 nm,
120 nm, 90 nm, 60 nm, 30
nm and 0 nm(contact). The curves are scaled to lie on top of one another by multiplying the frequency shift data by  a scale factor of $1.6\times10^{-12}$. 
\textbf{V. Parameter-free extraction of thin-film dielectric constants}

For thin film samples, manipulation of Eq. \ref{equ213} allows the scaling factor to be eliminated.  This approach proceeds by alternately measuring a thin-film sample and a bare substrate placed side by side on the ground plane.  For films that are only a few hundred
nanometers thick, the frequency shift is typically 20\% greater than for a bare substrate, especially
when the thin film's permittivity is close to the substrate's. The independent measurements give two sets of data:
\begin{equation}
    \frac{\Delta{f}}{f}_{thin film}=\frac{E_{pert-thin film}}{\alpha}
\end{equation}
\begin{equation}
    \frac{\Delta{f}}{f}_{substrate}=\frac{E_{pert-substrate}}{\alpha}.
\end{equation}
\\
Taking the ratio of these two expressions, $\alpha$ is canceled and film's
dielectric constant can be determined directly.  That is
\begin{equation}\label{equratio}
    \frac{\Delta{f}}{f}_{ratio}=\frac{\frac{\Delta{f}}{f}_{thin film}}{\frac{\Delta{f}}{f}_{substrate}} =\frac{E_{pert-thin
    film}}{E_{pert-substrate}}.
\end{equation}

\begin{figure}
  \centering
  \includegraphics[width=8cm]{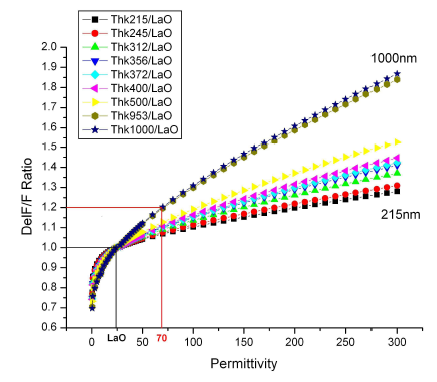}\\
  \caption{Computed $\Delta$f/f ratio curves, following Eq. \ref{equratio} for thin-film samples on a LaAlO$_3$ substrate. The sample thickness is varied by redrawing the calculation geometry, while $\epsilon$ is varied parametrically. The resulting curves are used to look up dielectric constants by finding the value of the permittivity which will give the measured $\Delta$f/f ratio. The example on the graph shows that a ratio of 1.2 would result from a film with a dielectric constant of 70.}\label{fig1:5}
\end{figure}

This approach is applied to a series of rare-earth, metal-oxide thin films grown on LaAlO$_3$, substrates
($\epsilon$=24) with different film thicknesses. By varying the thin film permittivity from 1 to 300 in the
finite-element calculation, reference curves for the perturbed energy \emph{vs.} dielectric constant with different film thicknesses are
calculated, as plotted in Fig. \ref{fig1:5}. Then, the experimental frequency shifts are converted to
thin-film dielectric constants by matching the measured value of the $\Delta{f}$/$f_{ratio}$ with the value
computed by the simulation.  For example, matching the experimental measurement of the frequency shift ratio (1.2 in this case) for a 1000 nm film, the corresponding $\epsilon$ is found to be 70, as shown in Fig. \ref{fig1:5}. To test the effect of the
substrate thickness, its value is varied from 500 $\mu$m to 2 mm in the calculation, resulting  in an immeasurable difference in frequency shift; thus the substrates are be treated as being infinitely thick in the calculation.

\textbf{VI. Determination of the imaginary part of the dielectric constant}

By a similar approach to the previous section, the imaginary part of the
permittivity is determined with the scanning near-field microwave microscope, a subject on which there are no reliable reports so far. Unlike the
simulation of real part of the permittivity, this model is based on a high-frequency EM model of Femlab for the
following reasons:
\begin{enumerate}

\item The loss in the
medium is due predominantly to dipolar motion, is intrinsically dynamic in nature, and thus the time-dependence of the electromagnetic field must be included in the calculation.
\item The real part of the permittivity can be obtained by solving Maxwell's equations over a
small volume, but the imaginary part needs to account for the radiation loss in the surrounding space.  That is, the perturbation of the electric field in the near-field region alone does not capture the entire loss mechanism, because the presence of a dielectric sample also changes the radiated field.
\end{enumerate}
The implementation is straightforward. A geometry similar to that used in the quasi-static model is used except
for its dimensions. Here the full wave equation is solved in the entire space below the cavity, and
the energy loss in the medium is obtained from the term (in Eq. \ref{equ216}) accounting for the size of the dielectric absorption of the EM field. In addition,
the resistive heat loss, the cavity loss, and radiation lost from the tip are included. Of these, only the cavity loss cannot be accurately computed (similar to the case for the static dielectric constant in the previous section), but this can be determined from a calibration against a
standard sample.

\begin{figure}
  \centering
  \includegraphics[width=8cm]{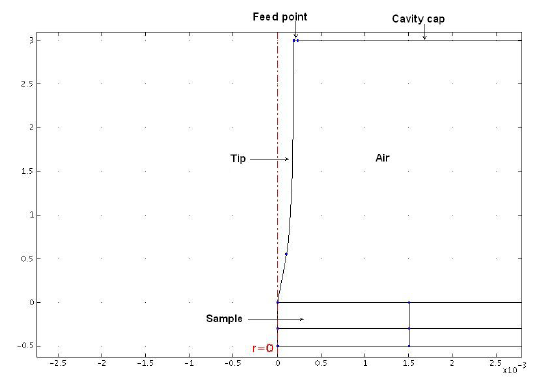}
  \caption{Geometry for the high frequency mode. The tip showing here is about 3mm long. The feed point is the gap between tip and sapphire disk, which connects the TEM field inside cavity with the radiation field outside. The difference between this and the static geometry (Fig. \ref{fig1:1} is that here the entire space between the cavity and the sample is discretized in order to compute the radiation losses.}\label{fig1:6}
\end{figure}

The calculation geometry, in Fig. \ref{fig1:6}, shows the space between the bottom of the coaxial cavity and the sample, where the electromagnetic field is primarily a TEM mode.  The feed point, which is the gap
between the tip and the hole in the cavity through which it protrudes, couples the EM field from inside to outside the cavity.  Part of this field travels along the wire extending from the cavity, down its length to the tip, where it is coupled to the sample as near-field radiation.  The remainder of the field dissipates radiatively into the free space between the sample and cavity cap.

The feed point is modeled to be a classical, first-order, low-reflecting boundary with an excitation field
source, H$_{\varphi0}$:

\begin{equation}
\vec{n}\times\sqrt{\epsilon}{\vec{E}}-\sqrt{\mu}\vec{H_{\varphi}}=-2\sqrt{\mu}\vec{H_{\varphi0}},
\end{equation}
where
\begin{equation}
H_{\varphi0}=\frac{Constant}{r}.
\end{equation}

\begin{figure}
  \centering
  \includegraphics[width=8cm]{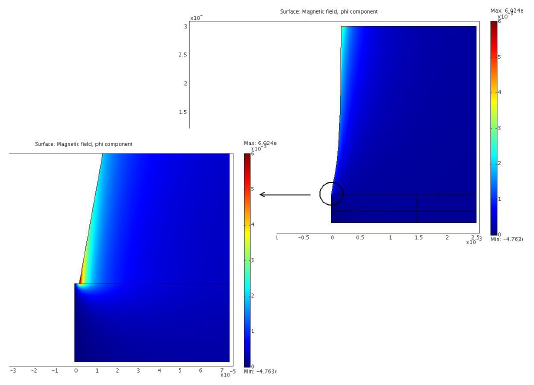}
  \caption{Solution for the electromagnetic field in high-frequency mode. The left-hand panel is a magnified view of the contact area. A strong electric field is present at the tip end, as is the case in quasi-static model.}\label{fig1:8}
\end{figure}

The tip, like any antenna, radiates into the free space and into the sample where a damped wave propagates. Since only a finite region can be discretized, the geometry is truncated ten centimeters from the tip using a similar absorbing
boundary condition as the feed point but without an excitation field.  In this model, the dielectric power dissipated is considered to be the total heat generated by the electromagnetic fields interacting
with the material: damping of the vibrating dipole moments and ohmic losses.  As is often done, both terms, $\omega\epsilon^{"}+\sigma$ are lumped together to be an effective conductivity
$\sigma_{eff}$. \cite{Pozar}A related quantity of interest is the loss tangent, defined to be
\begin{equation}
\tan\delta=\frac{\omega\epsilon^{"}+\sigma}{\omega\epsilon^{'}}.
\end{equation}
Thus, the absorption over the entire volume of the dielectric sample can be
expressed by:
\begin{equation}\label{equ215}
E_{resis.heat}=\frac{1}{2}\int\int\int_{v}Re[(\sigma_{eff}-j\omega\epsilon)\vec{E}\cdot\vec{E^{*}}].
\end{equation}
The solution proceeds by first drawing the geometry in the FemLab graphical environment and then generating a mesh. In areas where the field is likely to be intense, such as the
feed point, the tip surface, and the contact point with the sample, the mesh is refined to reduce the cell size to be as
small as 100 nm, for an accurate calculation. The solution obtained in this way is shown in Fig.
\ref{fig1:8}.

On the right side of Fig. \ref{fig1:8} is the full view of the simulated EM field, where we can see that the field
intensity is concentrated near the feed point, and it attenuates down the tip's length. In the open space
between the cavity cap and sample, the radiation field is weak but not insignificant, and the longer the tip, the greater the radiated power. This
is consistent with the experimental observation that longer tips will lower the cavity's quality factor, even though they are less well coupled to the sample.  Shorter tips have better coupling and lower loss, but the measurements are influenced more by the far-field
component propagating from the feed point, resulting in an increased sample-dependant background signal. In the setup reported here, a 3 mm long tip is used.

The left panel of Fig. \ref{fig1:8} is a close-up view of the field distribution at the tip end, where the field is seen to be concentrated at the contact point with the thin film, and to penetrate  only a few micrometers into the sample. Compared to the quasi-static model of the previous section, the accuracy of the near-field distribution in this model is lower, because the number of the mesh elements must be decreased in order to compensate for the increased computational cost of the time-dependent simulation. Nevertheless, the
essential characteristics of the near field are nearly the same as in the quasi-static model.

After solving for the electromagnetic field, the relationship between the quality factor and energy loss can be
established. Physically, the quality factor compares the stored energy of a resonant structure to the rate at
which it dissipates that energy.  For the experiments discussed here, the energy losses come from currents in the cavity walls, E$_{cavloss}$, radiation from the tip, E$_{tiploss}$, and ohmic heating in the sample, E$_{resis.heat}$. The corresponding quality factors
are:
\begin{equation}\label{equ216}
Q_{cav}=\omega\times\frac{E_{total}}{E_{cavloss}},
\end{equation}
\begin{equation}\label{equ217}
Q_{tip}=\omega\times\frac{E_{total}}{E_{tiploss}},
\end{equation}
\begin{equation}
Q_{resis.heat}=\omega\times\frac{E_{total}}{E_{resis.heat}},
\end{equation}
and the total Q is given by:
\begin{equation}\label{equ219}
\frac{1}{Q_{total}}=\frac{1}{Q_{cav}}+\frac{1}{Q_{tip}}+\frac{1}{Q_{resis.heat}}.
\end{equation}
Generally,  the average energy stored in cavity; E$_{total}$, and cavity wall loss, E$_{cavloss}$, are almost
impossible to measure, but these quantities can be determined from calibrations against known samples, in this
case, air, Q$_{air}$ and a standard, Q$_{sample}$. Since the majority energy in the system is stored inside
cavity, the sample is treated as a small perturbation, and E$_{total}$ can be assumed to be the same in both
measurements. E$_{cavloss}$ must also be the same because this factor is sample independent. Upon taking the
ratio of Q$_{air}$ to Q$_{sample}$, E$_{total}$ is canceled, leaving E$_{cavloss}$ as the only unknown factor. A
standard sample with a known loss tangent is measured to determine E$_{cavloss}$, and having determined this
parameter, other unknown samples can be measured. The calibration is performed against bulk LaAlO$_3$, with a 
loss tangent of $2\times10^{-5}$.\cite{Pozar,Christen96} The total effective conductivity is
then:
\begin{equation}
 \sigma_{eff} =tan\delta\times\omega\times\epsilon'=4.9\times10^{-5}.
\end{equation}

Without a sample, the measured resonant frequency is $\omega_{air}$=2$\pi$$\times$1756MHz, Q$_{air}$=878. With a
sample beneath the tip, the resonance frequency shifts to $\omega_{LaAlO_3}$=2$\pi$$\times$1747MHz,
Q$_{LaAlO_3}$=790. From Eqs. \ref{equ216} through \ref{equ219},
\begin{equation}
Q_{air}=\omega_{air}\times\frac{E_{total}}{E_{cavloss}+E_{tiploss-air}}
\end{equation}
\begin{eqnarray}
&&Q_{LaAlO_3}=\omega_{LaAlO_3}
\nonumber\\
&&\times\frac{E_{total}}{E_{cavloss}+E_{tiploss-LaAlO_3}+E_{resis.heat-LaAlO_3}}.
\end{eqnarray}
Here, E$_{tiploss}$, the power flow by radiation, is determined by integrating along the tip boundary using the
Femlab postprocessing function. These calculations show that when a sample is present, E$_{tiploss}$ increases
from $1.11\times10^{-9}$ to $2.33\times10^{-9}$. Similarly, E$_{resis.heat}$ is the sum of the resistive heating
over the entire thin film and substrate domains, which can be integrated by inserting $\sigma_{eff}$ into Eq.
\ref{equ215}. As mentioned previously, E$_{cavloss}$ is assumed to be independent of the sample and is the only
unknown parameter to be found by the calibration procedure.  Its value is found to be $19.132\times10^{-9}$,
which is almost 10 times greater than the sum of tip loss and resistive heating loss. By combining this value
with the the cavity's calibration factor, an unknown sample's loss tangent can be calculated. The results of
these calculations are listed in the Table \ref{tab1:1}.

\begin{table}[h]
\caption{\label{tab1:1}The size of each term in the total energy loss calculation. The resistive losses from the cavity walls dominate the others.}
\begin{tabular}{|l|l|l|}
  \hline
  Energy & Value & Source \\
  \hline
  E$_{total}$ & Unknown & Canceled  \\
  \hline
  E$_{tiploss-air}$ & $1.113\times10^{-9}$ & Femlab \\
  \hline
  E$_{cavloss}$ & $19.132\times10^{-9}$ & LaAlO$_3$ calib. \\
  \hline
  E$_{tiploss-LaAlO_3}$ & $2.331\times10^{-9}$ & Femlab \\
  \hline
  E$_{resis.heat-LaAlO_3}$ & $9.212\times10^{-10}$ & Femlab \\
  \hline
\end{tabular}
\end{table}

\begin{figure}
  \centering
  \includegraphics[width=8cm]{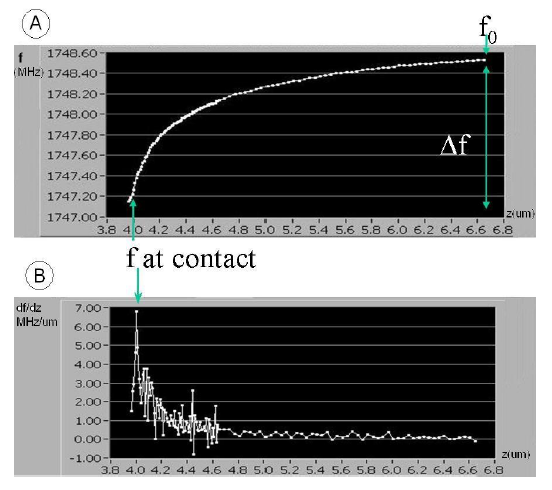}\\
  \caption{A plot of the cavity's frequency change \textit{vs.} tip-sample distance (Panel A), and its derivative (Panel B).  These measurements show a clear maximum in df/dz at the contact point.}\label{fig2:7}
\end{figure}

\textbf{VII. Experiment: Measurement of bulk and thin-film samples}
\\
Before measuring the dielectric constant of a sample, its thickness is measured and the copper stage is separated
from the tip by a distance exactly equal to the sample's thickness. At this position, the background resonant
frequency, f$_{0}$, is recorded, consistent with the boundary conditions set in the numerical simulation. The
sample is glued onto the stage with thermal cement and is brought steadily to within few microns of the tip.
\cite{Our00, YG01}  As the tip approaches, the resonant frequency begins to drop and the measurements begin when
the frequency drops to f$_{0}$.  When the sample is brought even closer to the tip, the frequency decreases more
dramatically, because of the near field coupling between the tip and the sample. Since accurate vertical
positioning is critical for our measurement, a piezoelectric actuator is used for the closest approach. Voltage
steps are applied to produce incremental movements as small as 30 nm, and the dielectric constant measurement is
made by stepping the sample towards the tip until the sample just contacts the tip. A typical plot of frequency vs. tip-sample separation
for a rare earth element thin film on a LaAlO$_3$ substrate is shown
in Fig. \ref{fig2:7}.

Panel A of Fig. \ref{fig2:7} is a plot of resonant frequency vs. approach distance. The distance is offset by the
sample position, which is determined by the tip-sample contact point.  The background frequency, f$_{0}$, as
mentioned above, is noted on graph to define the point at which the measurement begins.  The contact point is
determined by the df/dz curve, in Panel B, where as expected, the strongest interaction occurs and df/dz reaches
its maximum value. The piezo is allowed to advance 2 or 3 points past the contact point to insure
that peak is not caused by noise. Once the contact point is determined, the frequency shift, $\Delta{f}$, is
found, as shown on the graph. For low-dielectric-constant materials, the shift is small, and a typical
$\Delta{f}$/$f$ ratio is less than 1/1000.  The quality factor, Q, is obtained simultaneously, and Q$_{0}$ and Q$_{contact}$ 
are determined in the same way as f$_{0}$ and f$_{contact}$, with
the contact point determined from the df/dz plot. 
\begin{figure}
  \centering
  \includegraphics[width=8cm]{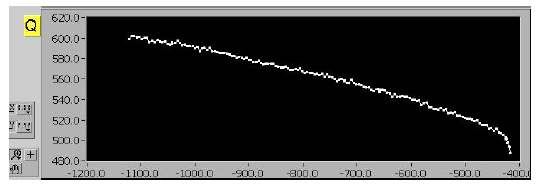}\\
  \caption{A plot of the quality factor \textit{vs.} tip-sample distance for a thin-film sample.  The offset on the horizontal axis is about 400 $\mu$.  Typically, the quality-factor data are more noisy than those for the frequency shift, and so the latter is used to determine the contact point.}\label{fig2:8}
\end{figure}

Using the measurement techniques and numerical modeling derived here, the dielectric
constant for both bulk and thin film samples are measured. Examples of measurements of a number of bulk materials are shown in
Table \ref{tab2:1}. The agreement between our measurements and the
literature values is good.

\begin{table}[h]
\caption{\label{tab2:1} Table of dielectric constant measurement for bulk samples.  All values are determined by
the same calibration constant from a TiO$_{2}$ sample.  Except for the teflon sample, the measured complex
dielectric constants agree well with accepted values. The TiO$_{2}$ value listed below is from a second sample,
measured to get a sense of the sample-to-sample variation in the dielectric constant measurement.}
\begin{tabular}{|l|l|l|l|l|}
  \hline 
  Sample & $\epsilon$ & $\epsilon$\cite{DavidRL} & tan$\delta$ & tan$\delta$\cite{KennethLK,Ivan} \\
  \hline 
     & Measured & Literature & Measured & Literature \\
  \hline
  Teflon & 3.16 & 2.4 & $3.6\times10^{-5}$ &   \\
  \hline
  Sapphire & 9.8 & 9.5 & $1.8\times10^{-5}$ & $1.4\times10^{-5}$ \\
  \hline
  MgO & 9.53 & 9.8 & $1.8\times10^{-5}$ & $1.6\times10^{-5}$ \\
  \hline
  STO & 278.2 & 276 & $1.804\times10^{-4}$ & $1.6\times10^{-4}$ \\
  \hline
  BTO & 292.23 & 300 & 0.513 & 0.47 \\
  \hline
  TiO$_{2}$ & 91.7 & 86 & $1.5\times10^{-2}$ & $1.2\times10^{-2}$ \\
  \hline
\end{tabular}
\end{table}

In Table \ref{tab4:1} are the permittivity data for a group thin films on LaAlO$_3$ substrates, determined by the
method of Eq. \ref{equratio}. The film's permittivities are shown in the last column. The consistency of the
approach is checked by comparing measurements of two pairs of films with identical compositions but with
significantly different film thicknesses, LaScO$_{3}$ and PrScO$_{3}$. Any errors resulting from changes in tip
geometry during the sequence of measurements or from numerical artifacts related to different film thicknesses
would be indicated by a disagreement in the values of $\epsilon$ for the samples of each
pair. As can be seen, excellent agreement is found, indicating the robustness of the measurement.\cite{Christen06}\\

\begin{table}[h]
\caption{\label{tab4:1} The measurement of rare-earth scandate thin films on LaAlO$_3$ substrates. By using the
ratio \textit{vs.} $\epsilon$ curves calculated from the simulation (Fig. \ref{fig1:5}), the films'
permittivities are found without introducing any fitting parameters.  The first and last two samples are noted
for having the same composition but different thicknesses. }
\begin{tabular}{|l|l|l|}
  \hline
  Thin film sample & Thickness(nm) & Film $\epsilon$ with error \\
  \hline
  \textbf{LaScO$_{3}$} & \textbf{953} & \textbf{32.3}\textbf{$\pm$}\textbf{1.5} \\
  \hline
  \textbf{PrScO$_{3}$} & \textbf{1000} & \textbf{29.6}\textbf{$\pm$}\textbf{1.2} \\
  \hline
  NdScO$_{3}$ & 309 & $47.0\pm3.0$ \\
  \hline
  SmScO$_{3}$ & 372 & $37.3\pm2.0$ \\
  \hline
  TbScO$_{3}$ & 312 & $38.7\pm4.7$ \\
  \hline
  GdScO$_{3}$ & 704 & $31.0\pm2.0$ \\
  \hline
  DyScO$_{3}$ & 215 & $31.3\pm2.1$ \\
  \hline
  HoScO$_{3}$ & 226 & $31.7\pm1.5$ \\
  \hline
  ErScO$_{3}$ & 245 & $19.7\pm1.5$ \\
  \hline
  YScO$_{3}$ & 500 & $20.3\pm1.5$ \\
  \hline
  \textbf{LaScO$_{3}$} & \textbf{400} & \textbf{32.3}\textbf{$\pm$}\textbf{1.2} \\
  \hline
  \textbf{PrScO$_{3}$} & \textbf{350} & \textbf{30.0}\textbf{$\pm$}\textbf{2.0} \\
  \hline
\end{tabular}
\end{table}

\textbf{VIII. Polarization, Sensitivity, and Spatial Resolution}

The sensitivity of the measurement is mainly determined by the tip flatness and tip-cone angle, both of which change as
the tip end is deformed during the measurements.
\\
\begin{figure}
  \centering
  \includegraphics[width=8cm]{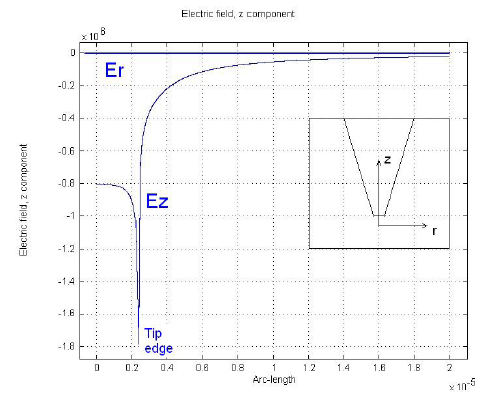}\\
  \caption{The electric field intensity for the radial and azimuthal polarizations near the tip end. Clearly, the azimuthal component is dominant, which makes the measurement primarily sensitive to the out-of-plane dielectric constant.}\label{fig1:9}
\end{figure}
\\
The spatial variation of the radial and azimuthal components of the electric field (See Fig. \ref{fig1:9}.) reveals much about the nature of the measurement.  First, the z-component is a few hundred times larger than r-component, which leads to the conclusion that the dominant component of the electric field is polarized along z-axis.  Second, the field reaches its maximum value at the edge of the tip, and then decays rapidly, which for this case defines E$_z$ by a 2.25 $\mu$m radius ring with 250 nm width, the actual interaction area of the sample with the electromagnetic field. This particular field distribution is qualitatively different from the field emanated by a spherical tip.\cite{YG01}
Overall the analysis indicates that the spatial resolution is almost identical to the diameter of the flat end of the tip, typically between 200 nm to 6 $\mu$m). Thus, for a high spatial resolution measurement, a small, sharp tip is must be used.

\begin{figure}
  \centering
  \includegraphics[width=8cm]{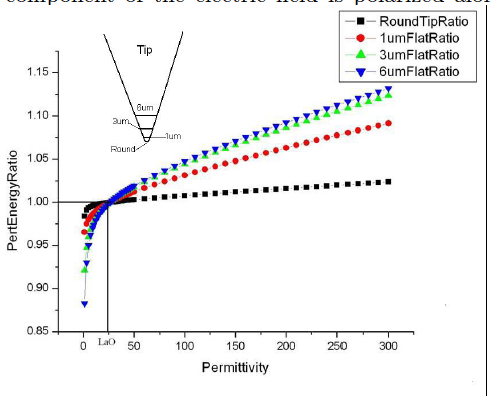}\\
  \caption{Sensitivity analysis of tips with different flatnesses. For a lanthanum aluminate substrate, the thin film's permittivity is varied from 1 to 300. Clearly, the flatter the tip, the more sensitive the permittivity measurement, until the flattening exceeds
  3$\mu$m, at which point, the sensitivity begins to saturate.}\label{fig1:10}
\end{figure}

The dielectric sensitivity is also influenced by the tip-cone angle and flatness, because these
determine the effective area of the tip and its coupling to the sample. The $\Delta{f}$/$f$ ratio
for different values of tip flatness is calculated, and the results are plotted in Fig. \ref{fig1:10} It is found that the flatter the tip, the higher the
sensitivity (shown by a larger change in perturbed energy), but when the flatness exceeds 3 $\mu$m, the
sensitivity saturates. Thus, Fig. \ref{fig1:9} can be used to optimize the tip geometry that achieves the tradeoff between high measurement sensitivity and the desired spatial resolution.\\

V. ACKNOWLEDGMENTS

The authors acknowledge David Norton for valuable discussions and for providing some of the bulk samples. This work
was supported by funding from the Army Research Office and the Office of Naval Research.

%

%
\pagebreak

%

%
%

%

%
\end{document}